\documentclass[]{aipproc}

\newcommand{\araa}{ARAA}
\newcommand{\apj}{ApJ}
\layoutstyle{6x9}

\begin{document}

\title{GRB Cosmology and the First Stars}

\classification{95.30.Lz; 97.10.Bt; 97.20.Wt; 98.70.Rz}
\keywords      {cosmology: theory -- early universe -- gamma rays: bursts}

\author{Volker Bromm}{
  address={Astronomy Department, University of Texas, Austin, TX  78712}
}

\author{Abraham Loeb}{address={Astronomy Department, Harvard University, Cambridge, MA 02138}}

\begin{abstract}
 
Gamma-Ray Bursts (GRBs) are unique probes of the cosmic star formation
history and the state of the intergalactic medium up to the redshifts of the
first stars. In particular, the ongoing {\it Swift} mission might be the
first observatory to detect individual Population~III stars, provided that
the massive, metal-free stars were able to trigger GRBs. {\it Swift} will
empirically constrain the redshift at which Population~III star formation
was terminated, thus providing crucial input to models of cosmic
reionization and metal enrichment.
 
 \end{abstract}

\maketitle

\section{Introduction}

Gamma-Ray Bursts (GRBs) are believed to originate in compact remnants
(neutron stars or black holes) of massive stars. Their high luminosities
make them detectable out to the edge of the visible universe \cite{CL00,LR00}.
GRBs offer the opportunity to detect
the most distant (and hence earliest) population of massive stars, the
so-called Population~III (Pop~III henceforth), one star at a time.  In the
hierarchical assembly process of halos which are dominated by cold dark
matter (CDM), the first galaxies should have had lower masses (and lower
stellar luminosities) than their low-redshift counterparts.  Consequently,
the characteristic luminosity of galaxies or quasars is expected to decline
with increasing redshift. GRB afterglows, which already produce a peak flux
comparable to that of quasars or starburst galaxies at $z\sim 1-2$, are
therefore expected to outshine any competing source at the highest
redshifts, when the first dwarf galaxies have formed in the universe.

The first-year polarization data from the {\it Wilkinson Microwave
Anisotropy Probe} ({\it WMAP}) indicates an optical depth to electron
scattering of $\sim 17\pm 4$\% after cosmological recombination \cite{Kog03,Spe03}.
This implies that the first stars must have
formed at a redshift $z\sim 20$, and reionized a substantial fraction of
the intergalactic hydrogen around that time \cite{C03,CFW03,SL03,WL03,YBH04}.
Early reionization can be achieved
with plausible star formation parameters in the standard $\Lambda$CDM
cosmology; in fact, the required optical depth can be achieved in a variety
of very different ionization histories since {\it WMAP} places only an
integral constraint on these histories \cite{HH03}. One would
like to probe the full history of reionization in order to disentangle the
properties and formation history of the stars that are responsible for
it. GRB afterglows offer the opportunity to detect stars as well as to
probe the ionization state \cite{BarL04} and metal enrichment
level \cite{FL03} of the intervening intergalactic medium
(IGM).

GRBs, the electromagnetically-brightest explosions in the universe, should
be detectable out to redshifts $z>10$ \cite{CL00,LR00}.
High-redshift GRBs can be identified through infrared
photometry, based on the Ly$\alpha$ break induced by absorption of their
spectrum at wavelengths below $1.216\, \mu {\rm m}\, [(1+z)/10]$. Follow-up
spectroscopy of high-redshift candidates can then be performed on a
10-meter-class telescope. Recently, the ongoing {\it Swift} mission
\cite{Geh04} has detected a GRB originating at $z\simeq 6.3$
(e.g., \cite{Hai05}), thus demonstrating the viability of
GRBs as probes of the early universe.

There are four main advantages of GRBs relative to traditional cosmic
sources such as quasars:

\noindent {\it (i)} The GRB afterglow flux at a given observed time lag
after the $\gamma$-ray trigger is not expected to fade significantly with
increasing redshift, since higher redshifts translate to earlier times in
the source frame, during which the afterglow is intrinsically brighter
\cite{CL00}. For standard afterglow lightcurves and spectra, the
increase in the luminosity distance with redshift is compensated by this
{\it cosmological time-stretching} effect.

\noindent {\it (ii)} As already mentioned, in the standard $\Lambda$CDM
cosmology, galaxies form hierarchically, starting from small masses and
increasing their average mass with cosmic time. Hence, the characteristic
mass of quasar black holes and the total stellar mass of a galaxy were
smaller at higher redshifts, making these sources intrinsically fainter
\cite{WL02}.  However, GRBs are believed to originate from a
stellar mass progenitor and so the intrinsic luminosity of their engine
should not depend on the mass of their host galaxy. GRB afterglows are
therefore expected to outshine their host galaxies by a factor that gets
larger with increasing redshift.

\noindent {\it (iii)} Since the progenitors of GRBs are believed to be
stellar, they likely originate in the most common star-forming galaxies at
a given redshift rather than in the most massive host galaxies, as is the
case for bright quasars \cite{BarL04}. Low-mass host galaxies
induce only a weak ionization effect on the surrounding IGM and do not
greatly perturb the Hubble flow around them. Hence, the Ly$\alpha$ damping
wing should be closer to the idealized unperturbed IGM case
and its detailed spectral shape should be easier
to interpret. Note also that unlike the case of a quasar, a GRB afterglow
can itself ionize at most $\sim 4\times 10^4 E_{51} M_\odot$ of hydrogen if
its UV energy is $E_{51}$ in units of $10^{51}$ ergs (based on the
available number of ionizing photons), and so it should have a negligible
cosmic effect on the surrounding IGM. 

\noindent
{\it (iv)} GRB afterglows have smooth (broken power-law) continuum spectra
unlike quasars which show strong spectral features (such as broad emission
lines or the so-called ``blue bump'') that complicate the extraction of IGM
absorption features. In particular, the continuum extrapolation into the
Ly$\alpha$ damping wing (the Gunn-Peterson absorption trough) during
the epoch of reionization is much more straightforward for the smooth UV
spectra of GRB afterglows than for quasars with an underlying broad
Ly$\alpha$ emission line \cite{BarL04}.

Although the nature of the central engine that powers the relativistic jets
of GRBs is still unknown, recent evidence indicates that long-duration GRBs
trace the formation of massive stars (e.g.,
\cite{T97,Wij98,BN00,Kul00,BKD02,Nat05}) and in particular that long-duration
GRBs are associated with Type Ib/c supernovae \cite{Sta03}. Since the first
stars in the universe are predicted to be predominantly massive
\cite{ABN02,BCL02,BLar04}, their death might give rise to large numbers of
GRBs at high redshifts.  In contrast to quasars of comparable brightness,
GRB afterglows are short-lived and release $\sim 10$ orders of magnitude
less energy into the surrounding IGM. Beyond the scale of their host
galaxy, they have a negligible effect on their cosmological
environment\footnote{Note, however, that feedback from a single GRB or
supernova on the gas confined within early dwarf galaxies could be
dramatic, since the binding energy of most galaxies at $z>10$ is lower than
$10^{51}~{\rm ergs}$ \cite{BarL01}.}. Consequently, they are ideal probes
of the IGM during the reionization epoch.  Their rest-frame UV spectra can
be used to probe the ionization state of the IGM through the spectral shape
of the Gunn-Peterson (Ly$\alpha$) absorption trough, or its metal
enrichment history through the intersection of enriched bubbles of
supernova (SN) ejecta from early galaxies \cite{FL03}.  Afterglows that are
unusually bright ($>10$mJy) at radio frequencies should also show a
detectable forest of 21~cm absorption lines due to enhanced HI column
densities in sheets, filaments, and collapsed minihalos within the IGM
\cite{FL02}.

Another advantage of GRB afterglows is that once they fade away, one may
search for their host galaxies. Hence, GRBs may serve as signposts of the
earliest dwarf galaxies that are otherwise too faint or rare on their own
for a dedicated search to find them. Detection of metal absorption
lines from the host galaxy in the afterglow spectrum, offers an unusual
opportunity to study the physical conditions (temperature, metallicity,
ionization state, and kinematics) in the interstellar medium of these
high-redshift galaxies.
A small fraction ($\sim 10$) of the GRB afterglows are expected to
originate at redshifts $z>5$ \cite{BL02,BL06}.  This subset of
afterglows can be selected photometrically using a small telescope, based
on the Ly$\alpha$ break at a wavelength of $1.216\, \mu {\rm m}\,
[(1+z)/10]$, caused by intergalactic HI absorption.  The challenge in the
upcoming years will be to follow-up on these candidates spectroscopically,
using a large (10-meter class) telescope.  GRB afterglows are likely to
revolutionize observational cosmology and replace traditional sources like
quasars, as probes of the IGM at $z>5$.  The near future promises to be
exciting for GRB astronomy as well as for studies of the high-redshift
universe.

\section{Important Open Questions}

Using GRBs to probe the high redshift universe has great promise.
In the following, we discuss some key open questions.

\subsection{Cosmic Star Formation at High Redshifts}

It is of great importance to constrain the Pop~III star formation mode, and
in particular to determine down to which redshift it continues to be
prominent. The extent of the Pop~III star formation will affect models of
the initial stages of reionization (e.g., \cite{WL03,CFW03,Sok04,YBH04,ABS06})
and metal enrichment (e.g., \cite{MBH03,FL03,FL05,Sch03,SSR04}),
and will determine whether planned surveys
will be able to effectively probe Pop~III stars (e.g., \cite{Sca05}).
The constraints on Pop~III star formation will also determine
whether the first stars could have contributed a significant fraction to
the cosmic near-IR background (e.g., \cite{SBK02,SF03,Kas05,MS05,DAK05}).

To constrain high-redshift star formation, one has to carry out a
two-step approach:

\noindent {\it (1)} {\it What is the signature of GRBs that originate
in metal-free, Pop~III progenitors?} Simply knowing that a given GRB came
from a high redshift is not sufficient to reach a definite conclusion as to
the nature of the progenitor. Pregalactic metal enrichment was likely quite
inhomogeneous, and we expect normal Pop~I and II stars to exist in galaxies
that were already metal-enriched at these high redshifts \cite{BL06}. Pop~III and
Pop~I/II star formation is thus predicted to have occurred concurrently at
$z > 5$. How is the predicted high mass-scale for Pop~III
stars reflected in the observational signature of the
resulting GRBs? Our preliminary results indicate that
circumburst densities are systematically higher in Pop~III
environments. GRB afterglows will then be much brighter than for
conventional GRBs. In addition, due to 
the systematically increased
progenitor masses, the Pop~III distribution may be biased toward
long-duration events.

\noindent {\it (2)}
The modelling of Pop~III cosmic star formation histories has a number of
free parameters, such as the star formation efficiency and the strength of
the chemical feedback. The latter refers to the timescale for, and spatial
extent of, the distribution of the first heavy elements that were produced
inside of Pop~III stars, and subsequently dispersed into the IGM by
supernova blast waves. Comparing with theoretical GRB redshift
distributions one can use the GRB redshift
distribution observed by {\it Swift} to calibrate the free model
parameters. In particular, one can use this strategy to measure the
redshift where Pop~III star formation terminates.

\begin{figure}[tp!]
\includegraphics[height=.3\textheight]{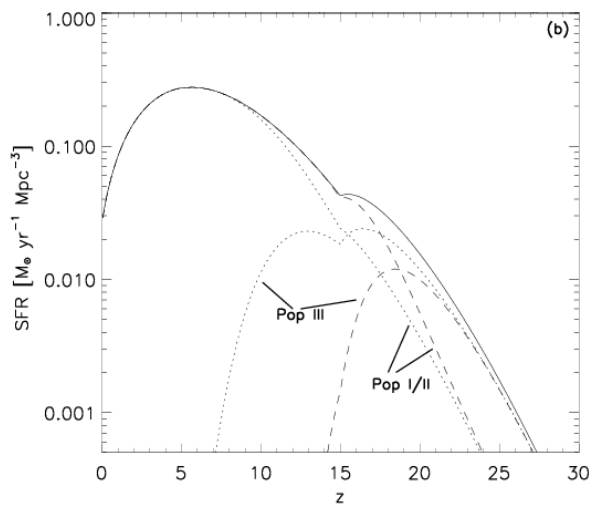}
\caption{Cosmic comoving star formation rate (SFR) in units of
$M_{\odot}$~yr$^{-1}$~Mpc$^{-3}$, as a function of redshift (from \cite{BL06}).
We assume
that cooling in primordial gas is due to atomic hydrogen only, a star
formation efficiency of $\eta_\ast=10\%$, and reionization beginning
at $z_{\rm reion}\approx 17$.  {\it Solid line:} Total comoving SFR.  {\it
Dotted lines:} Contribution to the total SFR from Pop~I/II and Pop~III for
the case of weak chemical feedback.  {\it Dashed lines:} Contribution to
the total SFR from Pop~I/II and Pop~III for the case of strong chemical
feedback.  
Pop~III star formation is restricted to high redshifts, but extends
over a significant range, $\Delta z\sim 10-15$.
\label{fig1}}
\end{figure}

In Figure~1 and 2, we illustrate this approach (based on \cite{BL06}).
Figure 2 leads to the robust expectation that $\sim 10$\% of all
{\it Swift} bursts should originate at $z > 5$. This prediction is based
on the contribution from Population~I/II stars which are known to exist
even at these high redshifts. Additional GRBs could be triggered by Pop~III
stars, with a highly uncertain efficiency. Assuming that long-duration GRBs
are produced by the collapsar mechanism, a Pop~III star with a close binary
companion provides a plausible GRB progenitor. We have estimated the
Pop~III GRB efficiency, reflecting the probability of forming sufficiently
close and massive binary systems, to lie between zero (if tight Pop~III
binaries do not exist) and $\sim 10$ times the empirically inferred value
for Population~I/II (due to the increased fraction of black hole forming
progenitors among the massive Population~III stars).

A key ingredient in determining the underlying star formation history from
the observed GRB redshift distribution is the GRB luminosity function,
which is only poorly constrained at present.  The improved statistics
provided by {\it Swift} will enable the construction of an empirical
luminosity function. With an improved luminosity function, we will be able
to re-calibrate the theoretical prediction in Figure~2 more reliably.

\begin{figure}[t]
\includegraphics[height=.3\textheight]{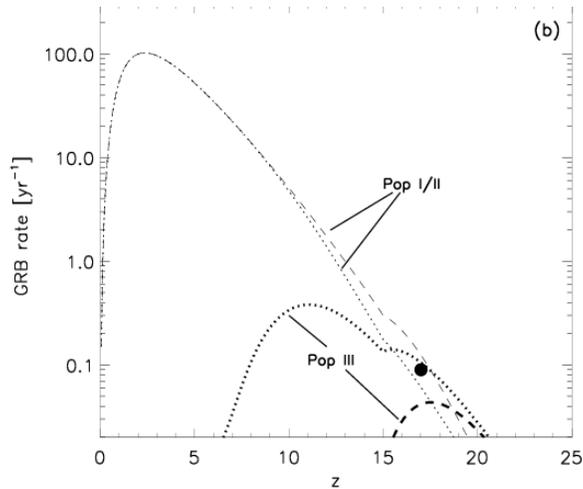}
\caption{Predicted GRB rate to be observed by {\it Swift} (from \cite{BL06}).
Shown is the observed
number of bursts per year, $dN_{\rm GRB}^{\rm obs}/d\ln (1+z)$, as a function
of redshift.  All rates are calculated with a constant GRB efficiency,
$\eta_{\rm GRB}\simeq 2\times 10^{-9}$~bursts $M_{\odot}^{-1}$, using the
cosmic SFRs from Fig.~1.
{\it Dotted lines:} Contribution to the observed GRB
rate from Pop~I/II and Pop~III for the case of weak chemical feedback.
{\it Dashed lines:} Contribution to the GRB rate from Pop~I/II and Pop~III
for the case of strong chemical feedback. 
{\it Filled circle:} GRB rate from Pop~III stars if these were
responsible for reionizing the universe at $z\sim 17$.  
\label{fig2}
}
\end{figure}

\subsection{Physical Properties of GRB Hosts}

In order to predict the observational signature of high-redshift GRBs, it
is important to know the properties of the GRB host systems.  
Within variants of the popular CDM model for structure
formation, where small objects form first and subsequently merge to build
up more massive ones, the first stars are predicted to form at $z\sim
20$--$30$ 
in minihalos of total mass (dark
matter plus gas) $\sim 10^6 M_{\odot}$
\cite{Teg97,BarL01,YBH04}.
These objects are the sites for the
formation of the first stars, and thus are the potential hosts of the
highest-redshift GRBs.
{\it What is the environment in which the earliest GRBs and their
afterglows did occur?}  
This problem breaks down
into two related questions: (i) what type of stars (in terms of mass,
metallicity, and clustering properties) will form in each minihalo?, and
(ii) how will the ionizing radiation from each star modify the density
structure of the surrounding gas? These two questions are fundamentally
intertwined. The ionizing photon production strongly depends on the stellar
mass, which in turn is determined by how the accretion flow onto the
growing protostar proceeds under the influence of this radiation field. In
other words, the assembly of the Population~III stars and the development
of an HII region around them proceed simultaneously, and affect each
other.  
As a first step (see Figure~3), 
we describe the photo-evaporation as a self-similar champagne flow \cite{Shu02}, 
with parameters appropriate for the Population~III case.

\begin{figure}[t]
\includegraphics[height=.3\textheight]{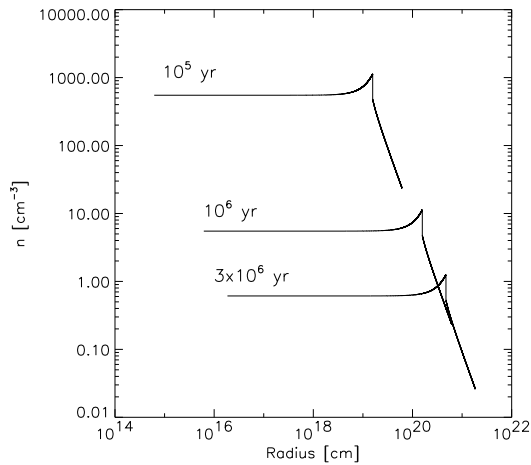}
\caption{Effect of photoheating from a Population~III star on the density
profile in a high-redshift minihalo.  The curves, labeled by the time after
the onset of the central point source, are calculated according to a
self-similar model for the expansion of an HII region. Numerical
simulations closely conform to this analytical behavior. Notice that the
central density is significantly reduced by the end of the life of a
massive star, and that a central core has developed where the density is
constant.}
\label{fig3}
\end{figure}

Notice that the central density is significantly reduced by the end of the
life of a massive star, and that a central core has developed where the
density is nearly constant. Such a flat density profile is markedly
different from that created by stellar winds ($\rho \propto
r^{-2}$). Winds, and consequently mass-loss, may not be important for
massive Population~III stars \cite{BHW01,K02},
and such a flat density profile may be
characteristic of GRBs that originate from metal-free Population~III
progenitors.

\begin{figure}[t]
\includegraphics[height=.3\textheight]{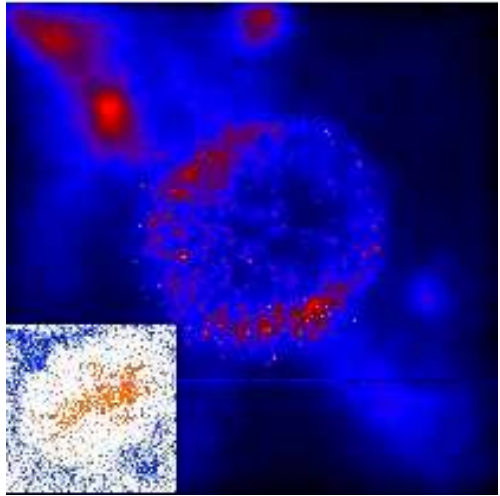}
\caption{Supernova explosion in the high-redshift universe (from \cite{BYH03}).
The snapshot is taken $\sim 10^6$~yr after the explosion with
total energy $E_{\rm SN}\simeq 10^{53}$~ergs. We show the projected gas
density within a box of linear size 1~kpc. The SN bubble has expanded to a
radius of $\sim 200$~pc, having evacuated most of the gas in the
minihalo. {\it Inset:} Distribution of metals. The stellar ejecta ({\it gray
dots}) trace the metals and are embedded in pristine metal-poor gas ({\it
black dots}).  }
\label{fig4}
\end{figure}

We have
carried out idealized simulations of the protostellar accretion
problem, allowing us to estimate the final mass of a Population~III star \cite{BL04}.
Using the smoothed particle hydrodynamics (SPH) method, we have included
the chemistry and cooling physics relevant for the evolution of metal-free
gas (see \cite{BCL02} for details). Improving on earlier work \cite{BCL99,BCL02}
by initializing our simulations according to the
$\Lambda$CDM model, we have focused on an isolated overdense region that
corresponds to a 3$\sigma-$peak \cite{BL04}: a halo containing a total mass of
$10^{6}M_{\odot}$, and collapsing at a redshift $z_{\rm vir}\simeq 20$.

We have found that one high-density clump has
formed at the center of the minihalo, possessing a gas mass of a few
hundred solar masses.  Soon after its formation, the clump becomes
gravitationally unstable and undergoes runaway collapse. Once the gas has
exceeded a threshold density of $10^{7}$ cm$^{-3}$, a sink particle is
inserted into the simulation.  This choice for the density threshold
ensures that the local Jeans mass is resolved throughout the simulation.
The clump (i.e., sink particle) has an initial mass of $M_{\rm Cl}\simeq
200M_{\odot}$, and grows subsequently by ongoing accretion of surrounding
gas.  High-density clumps with such masses result from the chemistry and
cooling rate of molecular hydrogen, H$_{2}$, which imprint characteristic
values of temperature, $T\sim 200$~K, and density, $n\sim 10^{4}$
cm$^{-3}$, into the metal-free gas \cite{BCL02}.  Evaluating the
Jeans mass for these characteristic values results in $M_{J}\sim \mbox{\ a
few \ }\times 10^{2}M_{\odot}$, which is close to the initial clump masses
found in the simulations.

The high-density clumps are clearly not stars yet. To probe the subsequent
fate of a clump, we have re-simulated the evolution of the central clump
with sufficient resolution to follow the collapse to higher densities (see
\cite{KW02,BL03} for a description of the refinement technique).  Our
refined simulation enables us to study the three-dimensional accretion flow
around the protostar (see also \cite{OP01,OP03,Rip02,TM04}).
We now allow the gas to reach densities of $10^{12}$
cm$^{-3}$ before being incorporated into a central sink particle. At these
high densities, three-body reactions \cite{PSS83}
have converted the gas into a fully molecular form.  We follow the growth
of the molecular core over the first $\sim 10^{4}$~yr after its formation,
making the idealized assumption that the protostellar radiation does not
affect the accretion flow.  The accretion rate is initially very high,
$\dot{M}_{\rm acc}\sim 0.1 M_{\odot}$~yr$^{-1}$, and subsequently declines
roughly as a power law of time. The mass of the molecular core, taken as a
crude estimate for the protostellar mass, grows approximately as:
$M_{\ast}\sim \int \dot{M}_{\rm acc}{\rm d}t \simeq 0.8
M_{\odot}(t/1\mbox{\ yr})^{0.45}$.  A robust upper limit for the final mass
of the star is then: $M_{\ast}(t=3\times 10^{6}{\rm yr})\sim 500
M_{\odot}$. In deriving this upper bound, we have conservatively assumed
that accretion cannot go on for longer than the total lifetime of a massive
star.

Our numerical results can be understood within the general theoretical
framework of how stars form \cite{L03}.  Star formation typically
proceeds from the `inside-out', through the accretion of gas onto a central
hydrostatic core.  Whereas the initial mass of the hydrostatic core is very
similar for primordial and present-day star formation \cite{ON98},
the accretion process -- ultimately responsible for setting the
final stellar mass -- is expected to be rather different. On dimensional
grounds, the accretion rate is simply related to the sound speed cubed over
Newton's constant (or equivalently given by the ratio of the Jeans mass and
the free-fall time): $\dot{M}_{\rm acc}\sim c_s^3/G \propto T^{3/2}$. A
simple comparison of the temperatures in present-day star forming regions
($T\sim 10$~K) with those in primordial ones ($T\sim 200-300$~K) already
indicates a difference in the accretion rate of more than two orders of
magnitude.

{\it Can a Population~III star ever reach this asymptotic mass limit?}  The
answer to this question is not yet known with any certainty, and it depends
on whether the accretion from a dust-free envelope is eventually terminated
by feedback from the star (e.g., \cite{OP01,OP03,Rip02,OI02,TM04}).
The standard
mechanism by which accretion may be terminated in metal-rich gas, namely
radiation pressure on dust grains \cite{WC87}, is
obviously not effective for gas with a primordial composition. Recently, it
has been speculated that accretion could instead be turned off through the
formation of an HII region \cite{OI02}, or through the
radiation pressure exerted by trapped Ly$\alpha$ photons \cite{TM04}.
The termination of the accretion process defines the current
unsolved frontier in studies of Population~III star formation. 

The first galaxies may be surrounded by a shell of highly enriched material
that was carried out in a SN-driven wind (see Fig.~4). A GRB in
that galaxy may show strong absorption lines at a velocity separation
associated with the wind velocity.  Simulating these winds and
calculating the absorption profile in the featureless spectrum of a GRB
afterglow, will allow us to use the observed spectra of high-$z$ GRBs
to directly probe the degree of metal enrichment in the vicinity of the
first star forming regions (see \cite{FL03} for a semi-analytic
treatment).

As the early afterglow radiation propagates through the interstellar
environment of the GRB, it will likely modify the gas properties close to
the source; these changes could in turn be noticed as time-dependent
spectral features in the spectrum of the afterglow and used to derive the
properties of the gas cloud (density, metal abundance, and size). 
The UV afterglow radiation can induce detectable
changes to the interstellar absorption features of the host galaxy \cite{PL98}; 
dust destruction could have occurred due to the GRB X-rays \cite{WD00,FKR01}, and
molecules could have been destroyed near the GRB source \cite{DH02}.
Quantitatively, all of the effects mentioned above strongly depend on the
exact properties of the gas in the host system.

\subsection{Population~III Progenitors}

Most studies to date have assumed a constant
efficiency of forming GRBs per unit mass of stars. This simplifying
assumption may either lead to an overestimation or underestimation of the
frequency of GRBs. Metal-free stars are thought to be massive \cite{ABN02,BCL02}
and their extended envelopes may suppress the
emergence of relativistic jets out of their surface (even if such jets are
produced through the collapse of their core to a spinning black hole). On
the other hand, low-metallicity stars are expected to have weak winds with
little angular momentum loss during their evolution, and so they may
preferentially yield rotating central configurations that make GRB jets
after core collapse. 

{\it What kind of metal-free, Pop~III
progenitor stars may lead to GRBs?} Long-duration GRBs appear to be
associated with Type Ib/c supernovae \cite{Sta03}, namely
progenitor massive stars that have lost their hydrogen envelope. This
requirement is explained theoretically in the collapsar model, in which
the relativistic jets produced by core collapse to a black hole are unable
to emerge relativistically out of the stellar surface if the hydrogen
envelope is retained \cite{MWH01}.  The question then arises as
to whether the lack of metal line-opacity that is essential for
radiation-driven winds in metal-rich stars, would make a Pop~III
star retain its hydrogen envelope, thus quenching any relativistic jets and
GRBs.

Aside from mass transfer in a binary system, individual Pop~III stars could
lose their hydrogen envelope due to either: (i) violent pulsations,
particularly in the mass range $100$--$140 M_{\odot}$, or (ii) a wind
driven by helium lines.  The outer stellar layers are in a state where
gravity only marginally exceeds radiation pressure due to
electron-scattering (Thomson) opacity. Adding the small, but still
non-negligible contribution from the bound-free opacity provided by
singly-ionized helium, may be able to unbind the atmospheric
gas. Therefore, mass-loss might occur even in the absence of dust, or any
heavy elements.

\end{document}